\documentclass[a4j,12pt,preprint]{jpsj2}
\usepackage{epsfig}

\def\runtitle{Basin structure in the two-dimensional dissipative circle map
}
\def\runauthor{
Yumino {\sc Hayase},
Shohei {\sc Fukano},
and Hiizu {\sc Nakanishi}
}

\title{Basin structure in the two-dimensional dissipative circle map}

\author
{
Yumino {\sc Hayase}\footnote{E-mail: yumino@stat.phys.kyushu-u.ac.jp},
Shohei {\sc Fukano}\footnote{E-mail: fukano@stat.phys.kyushu-u.ac.jp},
and Hiizu {\sc Nakanishi}\footnote{E-mail: naka4scp@mbox.nc.kyushu-u.ac.jp}
}

\inst{Department of Physics, Kyushu University 33, Fukuoka 812-8581}

\recdate{March 12, 2003}

\abst{Fractal basin structure in the two-dimensional dissipative circle
map is examined in detail.  Numerically obtained basin appears to be
riddling in the parameter region where two periodic orbits co-exist near
a boundary crisis, but it is shown to consist of layers of thin bands.}

\kword{fractal basin, riddled basin, circle map, transient chaos,
bouncing ball}

\makeatletter
\def\lsim{\mathrel{\mathpalette\gl@align<}}
\def\gsim{\mathrel{\mathpalette\gl@align>}}
\def\gl@align#1#2{\lower.6ex\vbox{\baselineskip\z@skip\lineskip\z@
    \ialign{$\m@th#1\hfil##\hfil$\crcr#2\crcr\sim\crcr}}}
\makeatother

\begin {document}
\sloppy
\maketitle
\section{Introduction}

The circle map is one of the model systems that have been studied
extensively in connection with the chaotic dynamics\cite{hgs}.  It
represents dynamics of various types of physical systems with
oscillatory motion under some external action, such as a kicked 
rotator, a bouncing ball \cite{everson, luna}, Josephson junctions in
microwave\cite{BBJ84,habip}, {\it etc.}  Depending upon the strength and
frequency of external field, it shows both quasi-periodic and mode-locked
motions;  Arnol'd tongue is a famous pattern that appears in the
bifurcation diagram for these behaviors.

Recently, the present authors showed that this map has peculiar basin
structure in the parameter region where two periodic orbits co-exist
near a boundary crisis\cite{fukano}.
Numerically obtained basin appears to consist
of a lot of dots and looks almost like a riddled basin, {\it i.e.} the
basin that is a closed set without interior, yet with finite measure.

The riddled basin has been shown to appear for a chaos attractor that is
confined within a subspace of the phase space\cite{pikovsky, ott}, but
in the case of periodic attractor, there exists a finite trapping
region, therefore, the basin should be an open set and cannot be a
riddled basin.

The riddled-like structure in this case has been found to come from long
transient chaotic motion and very small stability window for the
periodic attractor within the transient chaotic ``attractor''\cite{fukano}.

In this paper, we investigate in detail this riddled-like basin
structure in the two-dimensional dissipative circle map, and show the
basin actually consists of a lot of thin bands in spite of the
appearance of the numerically obtained basin.


\section{Two-dimensional circle map and riddled-like basin}

The two-dimensional dissipative circle map is the map for the two
variables $\theta$ and $r$,
\begin{eqnarray}
\theta_{n+1} &=&\theta_n + \Omega -\frac{K}{2 \pi} \sin (2 \pi \theta_n) + b
r_n \quad \mbox{mod 1},
\label{eq:2cir1}\\
r_{n+1} &=& b r_n - \frac{K}{2 \pi}\sin  (2 \pi \theta_n)
\label{eq:2cir2},
\end{eqnarray}
where $\Omega$, $K$, and $b$ are parameters, and the dynamics is
dissipative when $b<1$.  This map is equivalent with the bounce map that
describes the motion of bouncing ball on a vibrating platform within the
high bounce approximation\cite{com1}; The basin structure of the bounce
map was shown to be completely scattered and appear almost
riddling\cite{fukano}.

As for the 2-d circle map, the basin becomes riddled-like for $K=6$,
$\Omega=0.03138$, and $b=0.1$; With these parameters, there co-exist two
periodic attractors, namely, the period two and the period four
attractors.  In Fig.1, we show the basin structure of the period four
attractor by plotting dots at the points within the basin out of
$2000\times 2000$ grid points in the phase space, i.e. for each gid
points, we examine the orbits, and if the orbit ends up in the the
period four attractor, a dot is plotted at the point where we started,
but if the orbit ends up in the period two attractor, the point is left
blank.  We see a very much riddled-like basin although it cannot be a
riddled-basin in the mathematical definition\cite{fukano,pikovsky,ott}.

In the following, we investigate the detailed structure of such a basin.


\section{One-dimensional circle map}

First, we explore the case of $b = 0$, then
the variable $r$ becomes irrelevant for the dynamics, and
eq.(\ref{eq:2cir1}) reduces to the one-dimensional circle map
\begin{eqnarray}
\theta_{n+1} &=& f(\theta_n) \nonumber\\ 
             &=& \theta_n + \Omega -\frac{K}{2 \pi} \sin (2 \pi \theta_n)
\quad \mbox{mod 1}.
\label{eq:1cir}
\end{eqnarray}
It is well known that, for $|K| < 1$, one finds the Arnol'd tongues
where the motion is mode locked, and this becomes the complete devil's
staircase at $|K| = 1$, then chaotic and non-chaotic regions are densely
interwoven in the parameter space for $|K|>1$; the last region is the
one we will look at carefully.

Two periodic attractors can co-exist in some parameter regions for $|K|
> 1$.  The bifurcation diagram is displayed in Fig.2(a) upon changing
$\Omega$ with $K=6$.  It shows the period doubling and chaos as
increasing $\Omega$.  If we blow up the parameter region $0.0596 <
\Omega < 0.06$ (Fig.2(b)), we can see another periodic attractor appears
around $\Omega=0.0598$.  In this parameter region, the period one
attractor is stable, whereas a new period three attractor appears and
bifurcates into the period six as decreasing $\Omega$.  Upon decreasing
$\Omega$ further, this attractor bifurcates into a chaos attractor, then
it is destroyed suddenly through a boundary crisis of the period one
basin.

In the 1-d map, the co-existence of the two periodic orbits can be seen
easily by considering the $n$'th iterates of eq.(\ref{eq:1cir}), {\it
i.e.} $f^n(\theta)$.  The period $n$ attractor $\theta_i$ ($1\le i\le
n$) is given by
\begin{eqnarray}
\theta_i = f^n (\theta_i) .
\label{eq:n-map}
\end{eqnarray}
The function $f^1(\theta)$ and $f^3(\theta)$ are plotted in Fig.3 for
$K=6$ and $\Omega=0.0598$.  We see that the both of $f^1(\theta)$ and
$f^3(\theta)$ have the stable solution of eq.(\ref{eq:n-map}), which
shows eq.(\ref{eq:1cir}) has both the period one and the period three
attractor at this parameter.  The absolute value of the slope of
$f^3(\theta)$ is much larger than 1 for most of the value of $\theta$,
which implies that the period three attractor is stabilized only for a
very small parameter region.

Another consequence that comes from the steepness of $f^3(\theta)$ is
the fact that the period three attractor has an extremely small trapping
region, {\it i.e.} a connected region that any orbit inside converges to
the attractor without leaving the region.  The basin can be constructed
as the union of its pre-iterates, and the fact that $f^3(\theta)$
consists of a lot of nearly vertical lines means that the pre-iterates
are even smaller than their images and scattered all over the
$\theta$-axis, therefore, the basin structures of these attractors are
scattered and mixed each other.

We measure the uncertainty exponent\cite{ll}, or the final state sensitivity
$\alpha$, which is defined as the exponent for the probability
$f_d(\epsilon)$ that two points separated from each other by small
distance $\epsilon$ in the phase space are in different basins;
\begin{equation}
f_d(\epsilon) \sim \epsilon^{\alpha}.
\label{alpha}
\end{equation}
If $\alpha$ is small, the final state is difficult to predict from the
initial state, which means the two basins are intricate.  For $K=6$ and
$\Omega = 0.0598$, the estimated value of the uncertainty exponent is
$\alpha = 6 \times 10^{-2}$.  This means that the fractal dimension of
the basin boundary is close to 1, or the dimension of the phase space
$d$, because the fractal dimension $d_f$ is related to $\alpha$
through\cite{ll}
\begin{equation}
d_f = d - \alpha .
\end{equation}

It is conceivable that, for large $n$, $f^n(\theta)$ consists of lines
that are nearly vertical, thus if the period $n$ attractor is stable for
large $n$ and coexists with another periodic attractor, the uncertainty
exponent should be close to $0$.  Note that the parameter region of
such co-existence is extremely small in general.

\section{Two-dimensional circle map}

\subsection{the $b=0$ case}

We consider the basin structure for the $b=0$ case in the full
two-dimensional phase space of the circle map (\ref{eq:2cir1}) and
(\ref{eq:2cir2}), although the map is essentially one-dimensional in
this case; eq.(\ref{eq:2cir1}) becomes eq.(\ref{eq:1cir}), and the
variable $r$ simply follows $\theta$ through eq.(\ref{eq:2cir2}).  We
discuss its basin structure in the full phase space for $K=6$ and
$\Omega =0.0598$, where the period one and the period three attractors
co-exist.

From eq.(\ref{eq:2cir2}), the trajectory in the phase space is on the line
\begin{equation}
r_{n+1} =  - \frac{K}{2 \pi}\sin  (2 \pi \theta_n) .
\label{eq:transient}
\end{equation}
Namely, $r_{n+1}$ is determined only by $\theta_{n}$, which means that
the point $(\theta_n,r_n)$ moves along the line (\ref{eq:transient})
while the system shows the transient chaos.  After some iterations,
$(\theta_n,r_n)$ eventually falls into the trapping region of either the
period one or the period three attractor on the line.  Figure 4 shows
the ``attractor'' of the transient chaos with the period one and the
period three attractor for $K=6$, $\Omega=0.0598$, and $b=0$, where the
transient chaos attractor is simply a line.

It is easy to understand that the basin is made of a lot of vertical
bands as shown in Fig.5(a), because the variable $r$ does not affect the
dynamics.  The examined grid points are the same as the one in Fig.1,
and the grid points that end up in the period three attractor are marked
by black dots, whereas the points that lead to the period one attractor
are left blank.  We may not see the bands thinner than the grid spacing;
The finer grid we use, the more thin bands show up.  The ratio,
however, of black dots to all the grid points examined does not change
significantly.

\subsection{the $0\lsim b\ll 1$ case}

The vertical band basin structure for the $b=0$ case will be modified
when $b\neq 0$.  Fig.5(b) shows the basin of the period three attractor
for the very small value of $b = 10^{-5}$ with the same $\Omega$ and $K$
as in Fig.5(a).  In Fig.5(b), the basin seems to be made of a lot of
dots in addition to some vertical bands.  We believe, however, the dotted
structure is numerical artifact: the vertical stripes for $b=0$ should
be deformed only slightly to be the diagonal stripes for such a small
$b$.


In the following, we examine the procedure to construct the basin of the
period three attractor for $b\ll 1$ to convince ourselves that it has a
layered structure.

Let us start by noting that, from eq.(\ref{eq:2cir1}), the point that
will be mapped at $\theta_{n+1}=\theta^*$ with the arbitrary $r_{n+1}$
at ($n+1$)'th step is on the line
\begin{equation}
r_n = -\frac{1}{b}  \left(
       \theta_n + \Omega -\frac{K}{2\pi}\sin(2\pi\theta_n)- \theta^*
                    \right)
\label{eq:line}
\end{equation}
at the $n$'th step.

The trapping region of the period three attractor has a very narrow
width in the $\theta$-direction, but should be elongated in the
$r$-direction for $b\ll 1$.  Suppose $I_0$ is a part of the trapping
region with the size $\ell\times b$, where $\ell$ ($b$) is the length of the
region in the $\theta$- ($r$-)direction and we take $\ell\leq b$ (Fig.6).  The
transient chaos ``attractor'' is now a band with the width $O(b)$ as we
see from eq.(\ref{eq:2cir2}).  Let $I_{1}$ be the pre-iterate of $I_0$,
then $I_{1}$ is stretched along the line (\ref{eq:line}).  The width of
$I_{1}$ in the $\theta$-direction is also $O(b)$ because of the relation
\begin{equation}
\theta_{n} = \theta_{n+1}- r_{n+1} - \Omega,
\label{eq:haba1}
\end{equation}
which we have from eqs. (\ref{eq:2cir1}) and (\ref{eq:2cir2}).
The extension of $I_{1}$ along the $r$-direction is $O(1)$ because
\begin{equation}
r_{n} = \frac{1}{b} \left(r_{n+1} - \frac{K}{2\pi}\sin(2\pi\theta_n)\right)
\label{eq:haba2}
\end{equation}
due to eq.(\ref{eq:2cir2}).

Then, we now consider how $I_{2}$, i.e. the pre-iterate of $I_{1}$, is
distributed.  Let the part of $I_{1}$ along the line
($\ref{eq:transient}$) with the width $b$ be $I^*_{1}$ (Fig.6).  The
most part of the pre-iterate of $I_1$, namely $I_2$, is outside of the
$r\sim O(1)$ region when $b\ll 1$, and only the pre-iterate of the part
$I^*_1$ stays in the region of $r\sim O(1)$.  The region $I_2$ consists,
in turn, of the bands (\ref{eq:line}) with the value $\theta^*$ of the
corresponding part of $I^*_1$.  Repeating the above procedure, we
obtained the basin made of a lot of thin bands.

\section{Numerical analysis}

We now present numerical evidences of the band structure obtained in the
above discussion, and estimate the distribution function of the band
width.

In order to study the local basin structure around an arbitrary point
$(\theta_0, r_0)$ that is in the period three basin, we examine the
points along the circle with the radius $\epsilon$ centered at the
point.  The function $F(\phi; \epsilon)$ plotted in Fig.7 is the
function that takes the value of 1 if the point on the circle at the
angle $\phi$ around the central point $(\theta_0, r_0)$ is in the period
three basin, and zero otherwise; the angle $\phi$ is measured from the
$r$-direction.  This function should take the value of 1 around
$\phi\approx 0$ and $\pi$ if the central point $(\theta_0, r_0)$ is in
the basin of a vertical thin band of the width smaller than $\epsilon$.

Figure 7 shows  $F(\phi; \epsilon)$ for three different values of $\epsilon$
around the same point $(\theta_0,r_0)=(0.133,0.1)$ for the parameters
$K=6$ and $\Omega=0.0598$ with the very small value of $b=10^{-5}$.

When $\epsilon=10^{-3}$, $F(\phi;\epsilon)$ looks random and hardly
shows any signs of the layered structure (Fig.7(a)).  About 5\% of the
points on the circle are in the period three basin.  As for
$\epsilon=10^{-4}$, the band structure of the basin is beginning to show
in the plateaus around $\phi\approx 0$ and $\pi$ (Fig.7(b)).  For
$\epsilon=10^{-5}$, $F(\phi;\epsilon)=1$ for all value of $\phi$
(Fig.7(c)).  From these results, we can estimate the width of the basin
$l$ at $(\theta_0,r_0)=(0.133,0.1)$ as $l\approx 5\times 10^{-5}$.

Repeating the above procedure for a lot of points of $(\theta_0,r_0)$,
we can determine the probability $R(l)$ that a point in the period three
basin is in the band whose width is smaller than $l$.  This is related
to the probability density $P(l)$ to pick a band of period three basin
with the width $l$ by
\begin{equation}
R(l)=\int_0^{l} P(l') dl' .
\end{equation}

Numerical results of $R(l)$ are shown in Fig.8 in the log-log plot for
each basin of the three cases of $b=0$, $10^{-5}$, and $0.1$; the
difference between $R(l)$ for $b=0$ and that for $b=10^{-5}$ are
invisible.  We see that $R(l)$'s are very flat over many decades of $l$,
and if we fit the plots in Fig.8 by the straight lines, namely, the power 
\begin{eqnarray}
R(l) \propto l^{\beta},
\end{eqnarray}
with the exponent $\beta$, we obtain the same value of $\beta$ for the
two competing basins, $\beta = 0.06$ for both period one and period
three basin in the case of $b=0$ and $10^{-5}$, and $\beta=0.03$ for
both period two and period four basin in the case of $b=0.1$.  For
larger $b$, the exponent $\beta$ becomes large, which means there are
more thin bands.


In principle, the exponents can be different for the two competing
basins; $\beta_1$ for the basin 1 and $\beta_2$ for the basin 2.  If
$p_n$ is the probability that a phase point belongs to the basin $n$,
then the probability $f_d(\epsilon)$ that two points separated from each
other by small distance $\epsilon$ in the phase space are in different
basins is given by
\begin{equation}
f_d(\epsilon) \sim p_1 R_1(\epsilon) + p_2 R_2(\epsilon)
 \sim \epsilon^{\beta_1} + \epsilon^{\beta_2} 
\end{equation}
for $\epsilon\ll 1$, where $R_n(\epsilon)$ is the probability
that a point in the basin $n$ is on the band whose width is
narrower than $\epsilon$.  From the definition of $\alpha$
by (\ref{alpha}), we have 
\begin{equation}
\alpha=\min_n \beta_n ,
\end{equation}
which is confirmed by estimating $\alpha$ independently,
$\alpha=0.06$ for $b=0$ and $10^{-5}$ and 0.03 for $b=0.1$.

The two exponents $\beta_1$ and $\beta_2$ are same for all of our cases
as mentioned above.  There are, however, no geometrical reasons that
impose the equality, therefore, this should come from the mechanism that
the basin structure is generated, namely, the basins are mixed through
stretching and folding by a transient chaos, but the detailed analysis
remains to be done.

\section{Summary and discussions}

From the analysis of the map and the numerical evidences, we have
concluded that the basins for the periodic attractors have the layered
band structure in the circle map for the parameter region where the two
periodic attractors co-exist.  This is quite different from the
impression one might have from the numerically obtained basin (Fig.1).
Even a small value of $b$ like $10^{-5}$ changes the appearance of the
basin rather drastically from the obvious layered structure in the $b=0$
case (Fig.5).

By examining a neighboring region of each point,
we have estimated the distribution of the band width of the basin, and
shown that the distribution is hardly changed by the small value of
$b=10^{-5}$ (Fig.8); this is quite a contrast with the change in the
basin appearance mentioned above.

This apparent paradox is resolved when we plot the basin with the
slightly slanted grid for the case of $b=0$ and find the result is
nearly the same with that for $b=10^{-5}$ (Figs.5(b) and 9).  In the
case of $b=0$, the basin must consist of vertical bands, but if the
basin is examined numerically on the slanted grid points, the dotted
structure is obtained.  This is because the numerous thin bands that
escape from the straight grid points graze through the slanted grid
points.

The Lyapunov graph of the circle map has been studied in the literature
and shown the structure is intricate especially for the region of large
$K$\cite{hgs,BM98}.  This should come from the intricate basin structure
studied here; If one try to plot the Lyapunov graph with only one set of
initial values for the variable, one would get an intricate structure
upon changing the parameters of the map as they go through the region
where the basin structure is intricate.  This is because the basin that
the fixed initial point belongs to changes wildly even if the
global change of the basin structure is minute.  In these regions, the
system shows multi stability and the Lyapunov exponents that correspond
to two or more attractors are chosen almost randomly in the graph.


The basin structure in the similar situation has been studied by Dronov
and Ott\cite{DO00}.  They studied the case where the riddled basin is
destroyed by the de-stabilization of chaos attractor due to a periodic
orbit, and they found ``stalactite structure'' of basin.  There are some
similarities in the mechanism to our case, and their stalactite basin
may be compared with our layered band structure of basin, but the major
difference is that, in their case, the chaos attractor is confined in a
subspace of the phase space as is necessary to have a riddled basin
while the chaos is embedded in the full phase space in our case.


In summary, we investigated the basin structure of the two-dimensional
circle map.  From the analysis in the $b\ll 1$ case, in the parameter
region where the period $m$ attractor coexists with the period $n$
attractor,
%
%
the basin structure is apparently riddled-like, but is made of a lot of
thin bands.

\newpage

\begin{figure}
\begin{center}
\epsfig{width=7cm,file=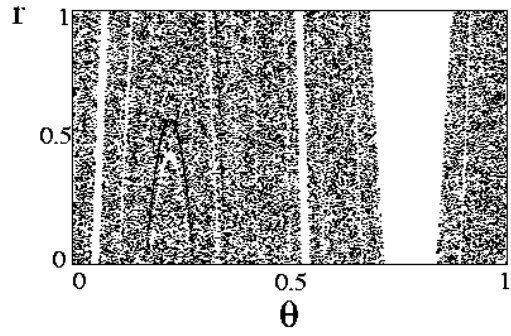}
\end{center}
\caption{The riddled-like basin of the period four attractor (dots) for
 the two-dimensional circle map with $K=6, \Omega=0.03138$ and $b=0.1$.}
\label{Fig-1}
\end{figure}

\begin{figure}
\begin{center}
\epsfig{width=7cm,file=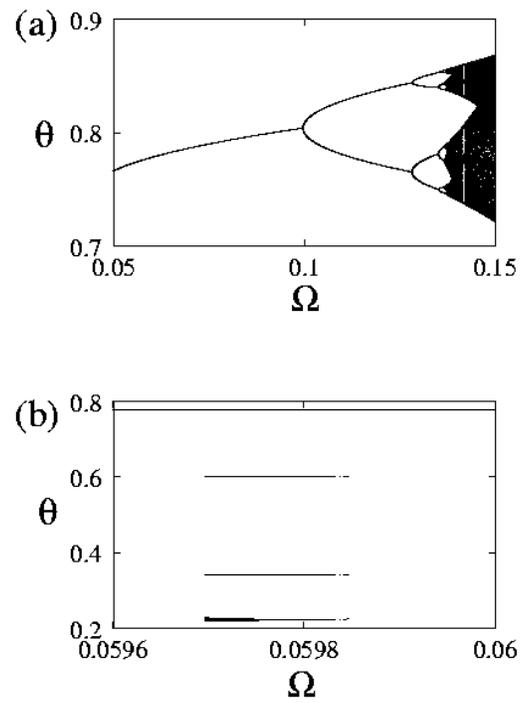}
\end{center}
\caption{Bifurcation Diagram for $b=0$ in the two-dimensional circle map
upon changing $\Omega$ with $K=6$.}
\label{Fig-2}
\end{figure}
\newpage
\begin{figure}
\begin{center}
\epsfig{width=12cm,file=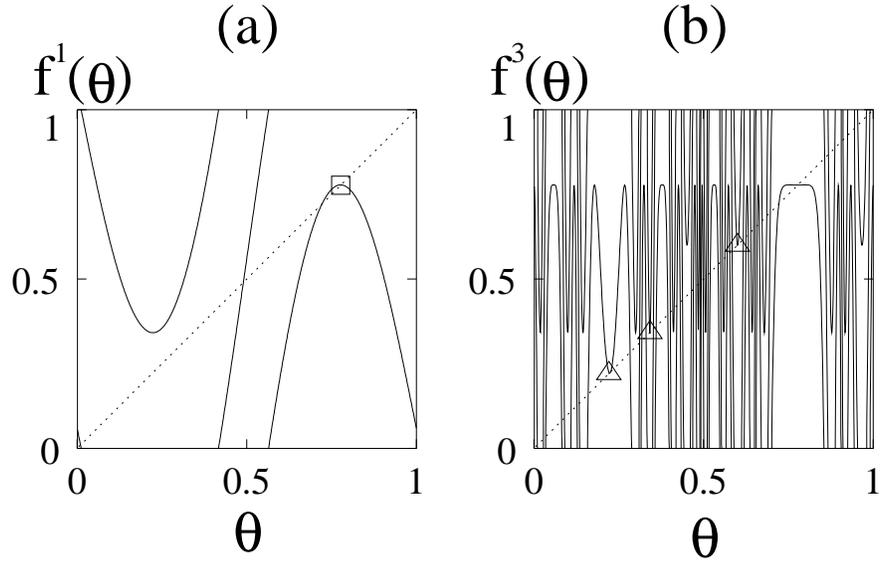}
\end{center}

\caption{The profile of $f(\theta)$ (a), and $f^3(\theta)$ (b) for $K=6$
and $\Omega=0.0598$ .  The dotted lines show $f(\theta) = \theta$. The
attractor of the period one ($\Box$) and the period three ($\triangle$)
are also indicated.}  \label{Fig-3}
\end{figure}
\begin{figure}
\begin{center}
\epsfig{width=7cm,file=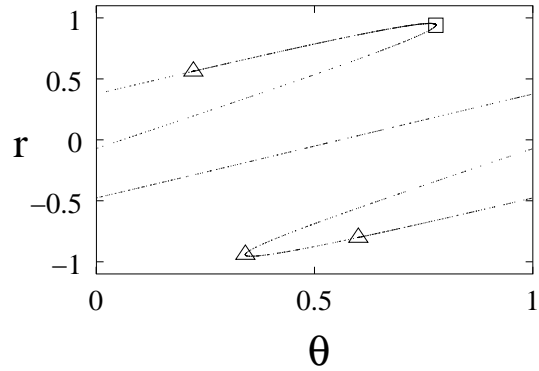}
\end{center}
\caption{The period three attractor ($\triangle$) and the
period one attractor ($\Box$) with the temporal attractor of the
transient chaos (dots) for $K=6, \Omega=0.0598$, and $b=0$.}
\label{Fig-4}
\end{figure}
\newpage
\begin{figure}
\begin{center}
\epsfig{width=6.5cm,file=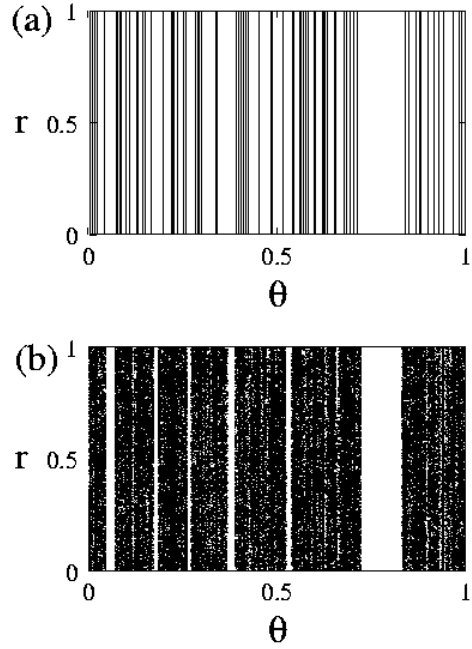}
\end{center}
\caption{The numerically obtained basin of the period three
attractor (dots) for $K=6, \Omega=0.0598$ for  $b=0$ (a) and  $b =
 10^{-5}$ (b).
The straight grid points of $2000\times 2000$ are examined.
}
\label{Fig-5}
\end{figure}
\begin{figure}
\begin{center}
\epsfig{width=5cm,file=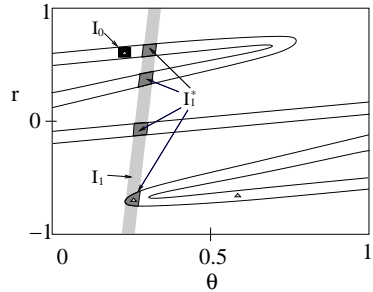}
\end{center}
\caption{ The schematic construction of the basin for the
period three attractor for $b\ll 1$.  The white band represents the
transient chaos ``attractor'' given by eq.(\ref{eq:transient}) with the
width $b$.  The region $I_0$ (black square) is the $\ell\times b$ ($\ell
<b$) region within the trapping region of the attractor.  The region
$I_1$ is the pre-iterate of $I_0$ and it extends along the line
(\ref{eq:line}).  The most of of the pre-iterate of $I_1$, or $I_2$,
goes out of the range shown in the graph; Only the region $I^*_1$, the
parts within the band, remain in the graph.  } \label{Fig-6}
\end{figure}
\newpage
\vskip .0cm
\begin{figure}
\begin{center}
\epsfig{width=8.5cm,file=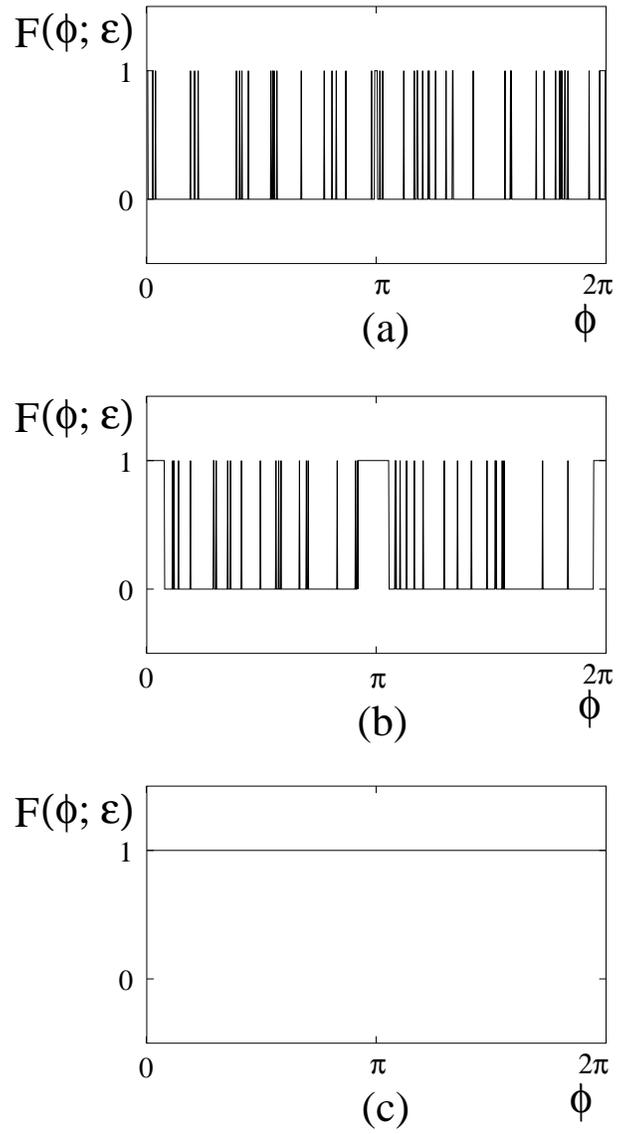}
\end{center}
\caption{ The function $F(\phi;\epsilon)$ around the point
$(\theta_0,r_0)=(0.133,0.1)$, which is in the period three basin, with
$K=6$, $\Omega=0.0598$, and $b=10^{-5}$ for $\epsilon=10^{-3}$ (a),
$10^{-4}$ (b), and $10^{-5}$ (c).  }  \label{Fig-7}
\end{figure}
\newpage
\begin{figure}
\begin{center}
\epsfig{width=6cm,file=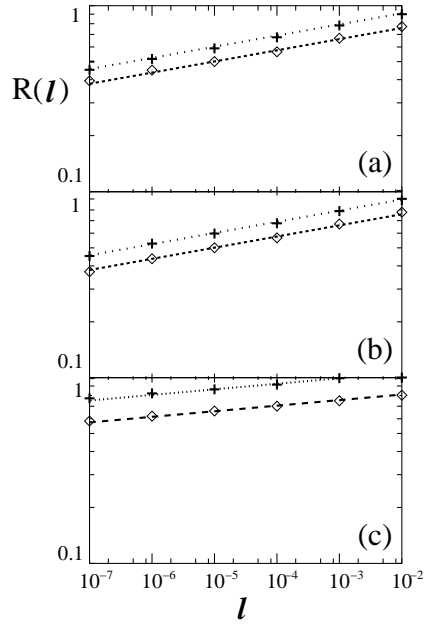}
\end{center}
\caption{ The probability $R(l)$ that the band width is
smaller than $l$ for each basin for $K=6$, $\Omega =0.0598$, and $b=0$
(a), $10^{-5}$ (b), and for $K=6$, $\Omega =0.03138$, and $b=0.1$ (c);
({\large$\diamond$}) for a majority basin (the period one or period two basin)
and ($+$) for a minority basin (the period three or period four basin).  The
dotted and dashed lines represent the lines of $l^\beta$ with $\beta=0.06$ for
$b=0$ and $10^{-5}$, and $\beta=0.03$ for $b=0.1$ } \label{Fig-8}
\end{figure}
\begin{figure}
\begin{center}
\epsfig{width=6cm,file=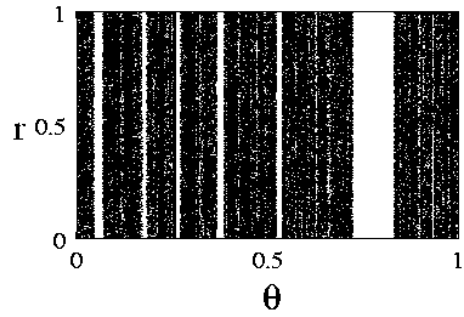}
\end{center}
\caption{The numerically obtained basin structure for $b=0$
using the slanted grids for $b=0$. The other parameters are same as in
Fig.5(a).}  \label{Fig-9}
\end{figure}

\end {document}